# Surface versus bulk state in topological insulator $Bi_2Se_3$ under environmental disorder

Matthew Brahlek[1], Yong Seung Kim[2], Namrata Bansal[3], Eliav Edrey[1], and Seongshik Oh[1,*]

[1]Department of Physics & Astronomy, Rutgers, The State University of New Jersey, 136 Frelinghuysen Rd, Piscataway, New Jersey 08854, U.S.A.

[2]Graphene Research Institute, Sejong University, Seoul 143-747, South Korea

[3]Department of Electrical and Computer Engineering, Rutgers, The State University of New Jersey, 94 Brett Rd., Piscataway, New Jersey 08854, U.S.A.

(Dated: June 08, 2011)

**Abstract:**

**Topological insulators (TIs) are predicted to be composed of an insulating bulk state along with conducting channels on the boundary of the material. In $Bi_2Se_3$, however, the Fermi level naturally resides in the conduction band due to intrinsic doping by selenium vacancies, leading to metallic bulk states. In such non-ideal TIs it is not well understood how the surface and bulk states behave under environmental disorder. In this letter, based on transport measurements of $Bi_2Se_3$ thin films, we show that the bulk states are sensitive to environmental disorder but the surface states remain robust.**

*Electronic mail: ohsean@physics.rutgers.edu

In the past decade there have been many advances in the class of materials known as topological insulators. It has been predicted that this class of materials exhibit a band structure similar to an ordinary insulator except the gap is spanned by metallic surface states (SS) with a linear, Dirac-like dispersion relation. The most remarkable character is that these SS remain robust against time reversal invariant disorder--robust in the sense that the SS are guaranteed to exist as long as time reversal symmetry is obeyed, although the properties that the carriers exhibit may fluctuate[1-3]. The first observations of SS were made in surface probes such as angle resolved photo-emission spectroscopy (ARPES) and scanning tunneling microscopy (STM), and more recently in direct transport measurements[4-10]. In this letter we report work done on $Bi_2Se_3$ thin films grown by molecular beam epitaxy, on Si (111) and on $Al_2O_3$ (0001) substrates (See Ref. 11 for more information on our growth procedure); all measurements were done at 1.5 K in a magnetic field up to 9T with exposure to air kept to less than 20 minutes between each measurement. We show how the transport properties changed during exposure to vacuum, water, pure gases as well as raw atmosphere; the SS remained robust after exposure to air while the bulk properties changed significantly.

It is known that band bending occurs at the surface of $Bi_2Se_3$[12]. This is seen by plotting the bottom of the conduction band (CB) versus depth from the surface (Fig. 1). After cleaving bulk samples, the bottom of the CB near the surface is above that of the bulk, forming a depletion layer (Fig. 1(a)). The material then relaxes to the thermodynamic ground state by natural n-type doping from either exposure to atmosphere and/or as Se atoms diffuse out, and over time the depletion layer becomes an n-type accumulation layer near the surface[7,12-15] (Fig. 1(b)). The exact time is highly dependent on the environment, and is estimated to be as little as a few seconds in atmosphere to as long as several hours in vacuum[12,14]. Despite the sensitivity of the Fermi level, ARPES done on bulk samples cleaved in atmosphere have shown that the SS still remains resolvable[14], suggesting that the SS survives the atmospheric disorders. However, in transport studies, the robustness of the SS predicted in theories and confirmed in ARPES measurements has never been confirmed. Instead, transport studies of bulk samples showed that the quantum oscillations of the SS, which were measured with the surface in the depletion state, disappears when the $Bi_2Se_3$ samples are exposed to air for an hour or so[7]. Although the latter experiment seems to be in contradiction to the robustness of the SS as observed in the ARPES measurement, in fact, it is not, because as the surface is exposed to air, the depletion layer transforms into an accumulation layer, which washes away the quantum oscillation signal even if the SS is still there. In other words, this previous transport study does not imply that the SS disappears with exposure to air, and whether the surface transport properties of $Bi_2Se_3$ survive atmospheric disorders is still an open question. Our recent achievement (Ref. 8) of isolating the surface transport properties even with the surface in the accumulation mode has made it possible in this letter to address this question.

Atmospheric gas is mainly composed of $N_2$, $O_2$, and water vapor. $N_2$ is basically chemically inert (a result of the strong triple bond between the N atoms), while $O_2$ readily ionizes. Therefore O and water are the most probable components to affect the electronic properties of $Bi_2Se_3$. The effect of O and water has to be taken in conjunction with the effects of the Se vacancies which n-dope the material (Fig. 2(a)). The rate at which each of these components diffuses in and out of the structure is a complex process, which depends both on the relative concentrations and on the intrinsic diffusion coefficient of each species[16]; due to the imprecise mixture of atmosphere, it is difficult to label what the ultimate effects on the electronic properties are.

To systemically test the effects of each component we exposed an 8 quintuple layer (QL—1 QL ≈ 1 nm) $Bi_2Se_3$ sample grown on Si to vacuum, and 1 atmosphere of $O_2$ and $N_2$ gas for periods of 1 week, and we used the standard Hall effect to determine the transport properties. Figure 2(b) shows the reversible tuning of the carrier density; when exposed to vacuum directly after growth the carrier density increased by 50%, and decreased by 50% when exposed to $O_2$. Finally when the sample was exposed to pure $N_2$ gas the carrier density increased by a similar amount; this implies that $N_2$ environment is almost equivalent to vacuum with regard to Se vacancy formation. The corresponding mobilities are also plotted in Fig. 2(b) on the right axis; the mobility measurements showed the same reversible effect confirming the origin to be the diffusion of Se and O in and out of the sample. The Se vacancies boost the carrier density and also act as point defects which decrease the mobility. O, which is chemically similar to Se at the impurity level, fills vacant Se slots, and thus pushes the mobility up and the carrier density down. The general trend in mobility is expected to be downward, because various contaminants will generally degrade the sample quality, and thus the fact that the mobility reversibly increased during oxygen exposure indicates that the sample quality remained nearly constant during this time.

Figure 2(c) shows data for another 8 QL sample on Si, which was exposed to vacuum for 2 weeks, followed by atmosphere for 2 days, and lastly liquid water for 1 hour. The carrier density in this sample similarly increased while in vacuum, dropped in atmosphere, but increased again in water. Data shown in Fig. 2(b)(c) suggest that $N_2$ (vacuum) and water provide n-type carriers through Se out-diffusion or water in-diffusion and only oxygen behaves as a p-type dopant; this observation is consistent with previous reports by other groups[7,15,17]. Accordingly, atmosphere, which is composed of $N_2$, $O_2$ and water, can function as either a p- or n-type dopant depending on which of these components dominates. In Fig. 2(c), two days of air exposure shows p-type doping, i.e. decrease in n-type carrier density, for the 8 QL sample, suggesting dominance of the oxygen effect. On the other hand, another sample (32 QL on Si) in Fig. 2(d) exhibited p-type doping initially, but after 20 hours, the trend changed to n-type doping, suggesting that initial dominance of the oxygen effect was later taken over by the selenium and water diffusion effects. These observations show that atmosphere can behave as both p- and n-type dopant depending on the detailed competition between oxygen and the rest.

Although the above study of films grown on Si substrates show that environment disorder has significant effects on the transport properties of $Bi_2Se_3$, they do not resolve the question of how the surface states are affected by these factors, because the carrier densities and mobilities in these samples were dominated by the bulk states[9]. As shown previously, samples grown on $Al_2O_3$ substrates enabled separation of the surface and bulk transport properties[8]. In the standard Hall effect, $R_{xy}$ scales linearly with the applied magnetic field. However, if two types of carriers contribute, each with a different mobility, say, for surface and bulk respectively, $R_{xy}$ scales nonlinearly with the applied magnetic field (see Fig. 3). For two types of carriers $n_1$ and $n_2$ with mobilities $\mu_1$ and $\mu_2$ respectively, the Hall resistance can be obtained by inverting the conductivity tensor with diagonal elements $G_{xx} = G_{yy}$, and off-diagonal elements $G_{xy} = -G_{yx}$, where

$$G_{xx}(B) = e\left(\frac{n_1\mu_1}{1+\mu_1^2 B^2} + \frac{n_2\mu_2}{1+\mu_2^2 B^2}\right); G_{xy}(B) = eB\left(\frac{n_1\mu_1^2}{1+\mu_1^2 B^2} + \frac{n_2\mu_2^2}{1+\mu_2^2 B^2}\right).$$

Upon inversion, the off-diagonal element of the resistance tensor, the Hall resistance, is found to be [8,18-19]

$$R_{xy}(B) = -\left(\frac{B}{e}\right)\frac{(n_1\mu_1^2 + n_2\mu_2^2) + B^2\mu_1^2\mu_2^2(n_1+n_2)}{(n_1\mu_1 + n_2\mu_2)^2 + B^2\mu_1^2\mu_2^2(n_1+n_2)^2}.$$

Using this function (in addition to $R_{xx}(B = 0)$, and $R_{xy}(B \to 0)/B$ which reduces the 4 parameters down to 2) we could fit the measured data and thereby determine the transport properties of the surface and bulk states[8]. In our previous work, we showed that over a range of 2-2750 nm in thickness one carrier density was independent of thickness while the other varied with the thickness[8]. From this observation we identified the first type of carriers with the surface layer, confined to 1 QL, and the second with the bulk. This allowed us to fix the thickness of our samples to determine how each of the surface and bulk transport properties changes under atmosphere.

Figure 3(a) and Table I show the measured data for $R_{xy}(B)$ for two 16 QL samples grown on $Al_2O_3$. The transport properties for sample A were measured directly after growth and measured again 2 days later. The dramatic change in character of the curves can be associated with an increase in bulk carriers. The fitting revealed that the bulk carrier density of sample A increased by 300% from $0.30\times10^{13}$cm$^{-2}$ to $1.1\times10^{13}$cm$^{-2}$. However, the surface carriers remained nearly constant during the exposure to atmosphere, changing only 10% from $3.3\times10^{13}$cm$^{-2}$ to $3.0\times10^{13}$cm$^{-2}$. The data obtained from sample B, which was grown under identical conditions and tested directly after growth, exhibited almost the same surface carrier density of $3.0\times10^{13}$cm$^{-2}$. However the bulk carrier density was $0.95\times10^{13}$cm$^{-2}$, which is over three times larger than that of sample A after growth. Regarding the mobilities, the bulk mobility of sample A showed a sharp decrease from 6400 to 1500 cm$^2$/Vs over two days. This drop is due to Se vacancies being produced along with contamination by other atmospheric impurities, both acting as scattering centers. In contrast, the surface mobility changed much less from 560 to 440 cm$^2$/Vs. These observations suggest that most of the changes observed in the samples grown on Si substrates are indeed from the bulk transport properties and in contrast, the surface states remain robust against air exposure. The robustness of the surface states in air is also supported by a recent optical measurement that observed strong surface transport on Bi$_2$Se$_3$ films left in air for several weeks[20].

To conclude, we investigated how the electronic properties of the topological insulator Bi$_2$Se$_3$ change under a variety of conditions. While the bulk transport properties were significantly affected by environmental factors such as vacuum, oxygen and moisture, the surface states remained stable. These observations resolve the seemingly previous discrepancy between transport and ARPES studies regarding the robustness of the surface states, and support the theoretical prediction of the surface state immunity to non-magnetic disorders. However, they raise another issue of the fragility of the bulk properties. Finding a way to stabilize the TI bulk properties seems to be the key to the future of TI applications.

We thank Peter Armitage, N. P. Ong and Johnpierre Paglione for useful discussion. This work is supported by IAMDN of Rutgers University, National Science Foundation (NSF DMR-0845464) and Office of Naval Research (ONR N000140910749).

**Figure captions**

**Figure 1** The dotted red curves show band bending tracking the bottom of the conduction band and the top of the valence band (VB) as a function of depth, x, from the surface of the material. The blue line, labeled as $E_f$, represents the Fermi energy. **(a)** Cleaved bulk samples exhibit an initial depletion region near the surface. **(b)** Unavoidable n-type doping occurs by exposure to atmosphere or vacuum, and the depletion region becomes an accumulations region. Thin films that are grown in vacuum and exposed to atmosphere always results in an accumulation region as in **(b)**.

**Figure 2 (a)** A cartoon showing the bulk contamination process. Se atoms diffuse out leaving Se vacancies (V), and O atoms and water molecules (W) naturally diffuse into the structure. **(b)** The carrier density (left axis open triangles), and the mobility (right axis solid squares); immediately after growth, 1 week in vacuum (Vac), 1 week in $O_2$ gas, and 1 week in $N_2$ gas. **(c)** A similar plot showing the response of the carrier density and mobility to exposure to vacuum for 2 weeks, air for 2 days, and lastly liquid water for 1 hour. **(d)** The response of the carrier density to air exposure.

**Figure 3 (a)** Non-linear Hall effect shows that two types of carriers contribute to the conduction. Sample A (solid squares) at t = 0 days, shows the lowest bulk carrier density; at t = 2 days (solid circles) the bulk carrier density of sample A has increased due to n-type doping, there-by overwhelming the surface signal and thus reducing the nonlinearity. The solid curve shows data from sample B whose bulk carrier density is initially larger. **(b)** The fitting parameters were extracted using $G_{xy}$ rather than $R_{xy}$ for functional simplicity. The fitted curves are represented by solid curves, while the experimental data are depicted by open triangles (sample A, t = 0), and open squares (sample B, t = 0).

**Table I** Taken from two 16 QL samples grown on $Al_2O_3$, this table shows the carrier density, $n_{SS}$ (surface state) and $n_{BS}$ (bulk state) and the corresponding mobilites obtained by the non-linear fitting process for $G_{xy}$.

**Fig. 1**

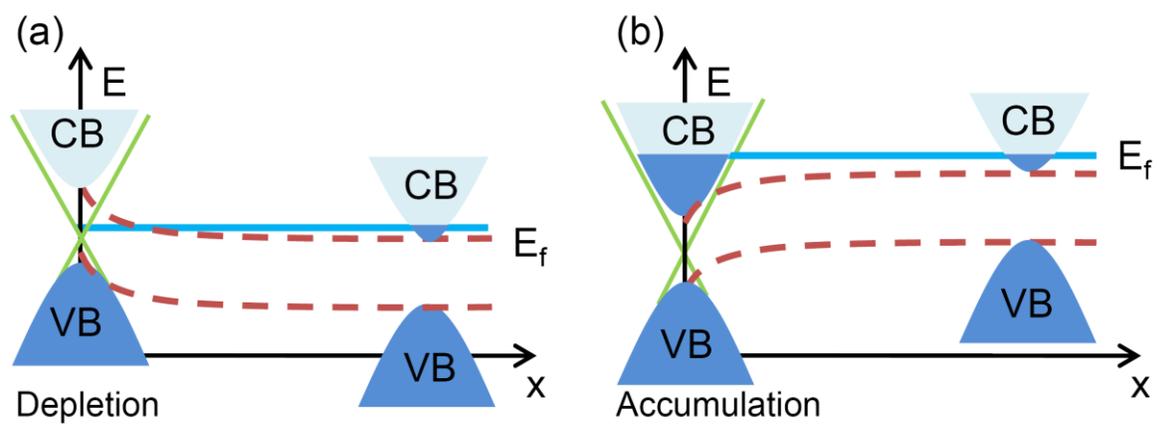

**Fig. 2**

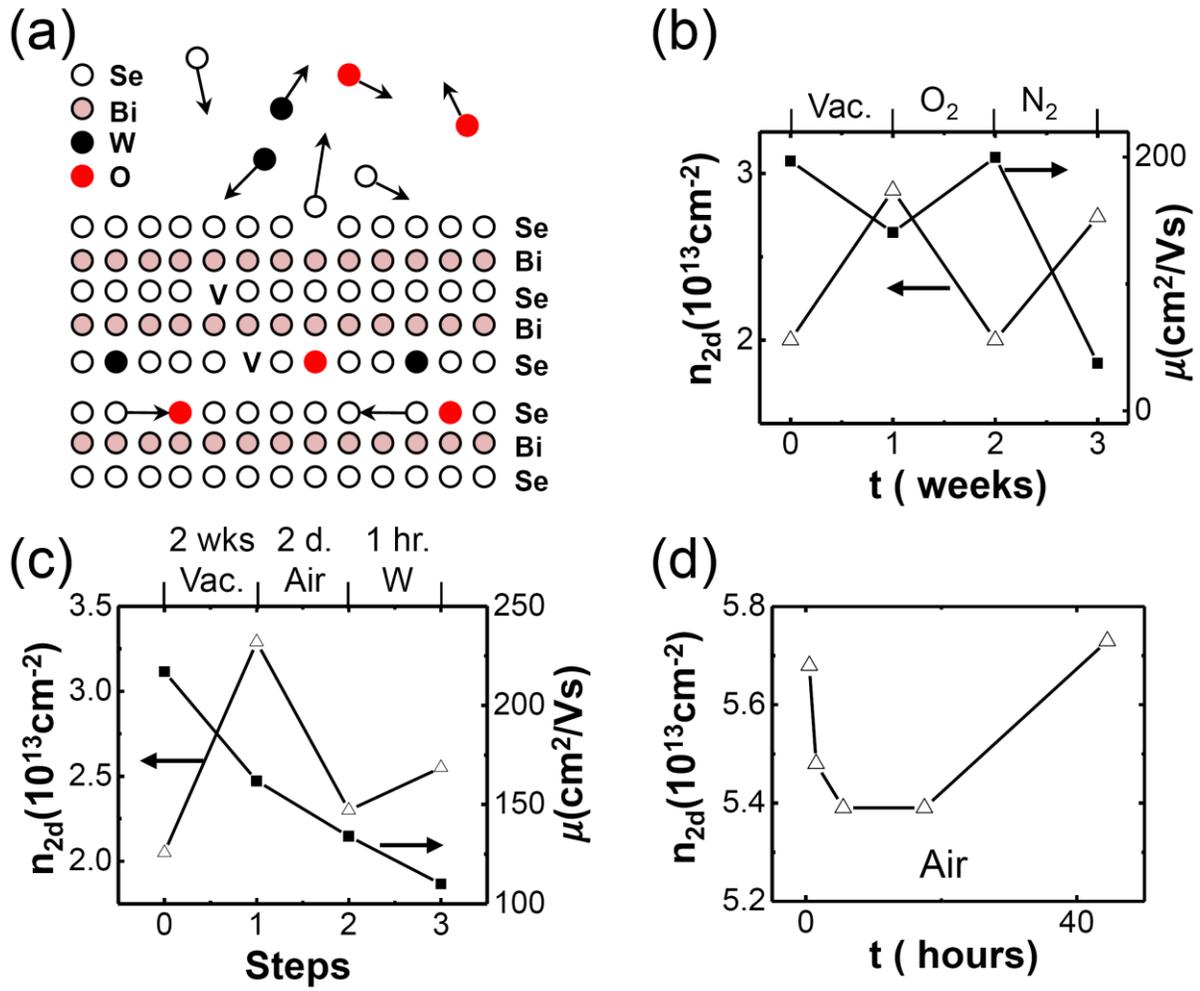

**Fig. 3**

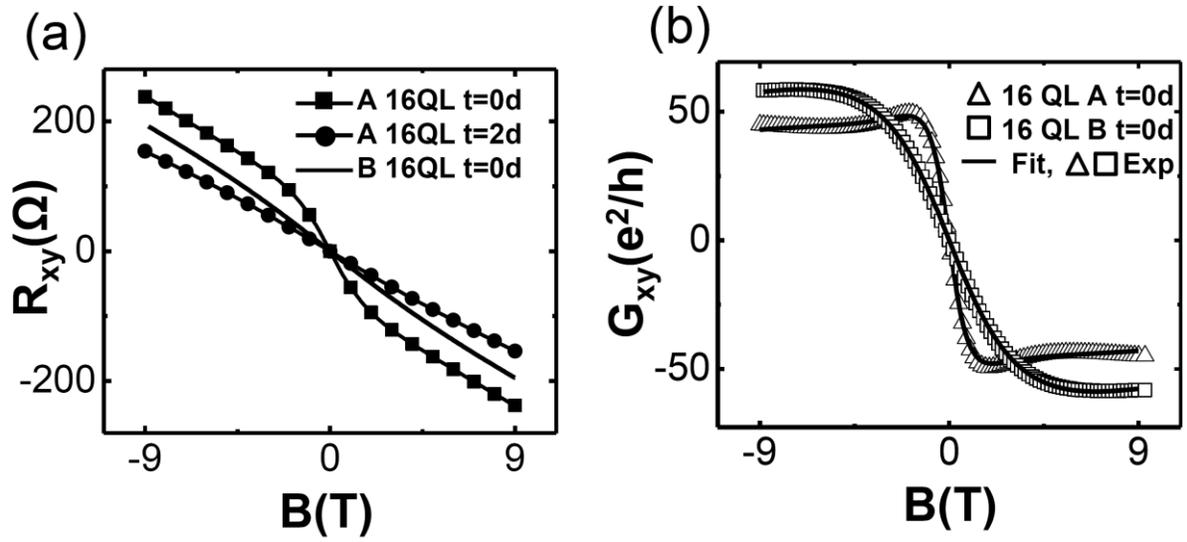

**Table I**

|  | $n_{SS}$ ($10^{13}$cm$^{-2}$) | $n_{BS}$ ($10^{13}$cm$^{-2}$) | $\mu_{SS}$ (cm$^2$/Vs) | $\mu_{BS}$ (cm$^2$/Vs) |
|---|---|---|---|---|
| A (t = 0.5 hrs) | 3.3 | 0.30 | 560 | 6400 |
| A (t = 48 hrs) | 3.0 | 1.1 | 440 | 1500 |
| B (t = 0.5 hrs) | 3.0 | 0.95 | 510 | 2100 |